# Denoising of Three-Dimensional Fast Spin Echo Magnetic Resonance Images of Knee Joints using Spatial-Variant Noise-Relevant Residual Learning of Convolution Neural Network

Shutian Zhao[1], Dónal G. Cahill[1], Li Siyue[1], Xiao Fan[1], Thierry Blu[2], James F Griffith[1], Weitian Chen[1]

1. Department of Imaging and Interventional Radiology, the Chinese University of Hong Kong

2. Department of Electrical Engineering, the Chinese University of Hong Kong

ABBREVIATIONS USED: 3D, three-dimensional; FSE, Fast Spin Echo; NEX, number of excitation; CNN, convolutional neural network; RL, residual learning; SNR, signal-to-noise ratio; PSNR, peak-signal-to-noise ratio; SSIM, structural similarity index.

**Abstract**

**Purpose**

Two-dimensional (2D) fast spin echo (FSE) techniques play a central role in the clinical magnetic resonance imaging (MRI) of knee joints. Moreover, three-dimensional (3D) FSE provides high-isotropic-resolution magnetic resonance (MR) images of knee joints, but it has a reduced signal-to-noise ratio compared to 2D FSE. Deep-learning denoising methods are a promising approach for denoising MR images, but they are often trained using synthetic noise due to challenges in obtaining true noise distributions for MR images. In this study, inherent true noise information from 2-NEX acquisition was used to develop a deep-learning model based on residual learning of convolutional neural network (CNN), and this model was used to suppress the noise in 3D FSE MR images of knee joints.

**Methods**

A deep learning-based denoising method was developed. The proposed CNN used two-step residual learning over parallel transporting and residual blocks and was designed to comprehensively learn real noise features from 2-NEX training data.

**Results**

The results of an ablation study validated the network design. The new method achieved improved denoising performance of 3D FSE knee MR images compared with current state-of-the-art methods, based on the peak signal-to-noise ratio and structural similarity index measure. The improved image quality after denoising using the new method was verified by radiological evaluation.

**Conclusion**

A deep CNN using the inherent spatial-varying noise information in 2-NEX acquisitions was developed. This method showed promise for clinical MRI assessments of the knee, and has potential applications for the assessment of other anatomical structures.

# 1. INTRODUCTION

The knee is the largest joint in the human body, with a complex and frequently injured anatomical structure.[1] It is a synovial hinge joint, and its stability is maintained by ligaments, tendons, a joint capsule, and menisci. Magnetic resonance imaging (MRI) is the most prominent non-invasive diagnostic modality, and its excellent visualization of soft tissue characteristics means it provides optimal imaging of the knee joint.[2] Fast spin echo/turbo spin echo (FSE/TSE)[3], the commercial implementation of rapid acquisition with relaxation enhancement (RARE)[4], plays a central role in clinical knee MRI. Current knee MRI protocols often consist of two-dimensional (2D) FSE sequences repeated in multiple planes to observe overall anatomical structure. However, this process is time-consuming and its low through-plane resolution results in partial volume effects. Three-dimensional (3D) FSE generates thinner sections and thus reduced partial volume effects, and may also be acquired with isotropic resolution and reformatted into an arbitrary plane to visualize complex anatomical structures. The application of 3D isotropic FSE in clinical practice may therefore markedly decrease the total time required for MR imaging examination.[5]

Commercial 3D FSE sequences typically use long echo trains with variable flip angles to achieve a reasonable scan time without excessive blurring.[6] As short T2 tissues are dominant in knee joints, image blurring often remains in 3D FSE MR images of the knee if long echo trains are used. The approaches used to reduce image blurring without increasing scan time, such as reducing the minimum flip angle of the echo train, are often accompanied by a reduced signal-to-noise ratio (SNR).[7] In addition, high isotropic resolution in 3D images is often obtained at the cost of the SNR. Therefore, denoising is desirable when 3D FSE is used for clinical MR imaging of the knee.

Traditional denoising methods have been commonly used for decades and include bilateral filtering[8], total variation (TV)-based regularization[9], nonlocal means (NLM)[10], and K-singular value decomposition (K-SVD)[11]. Block-matching 3D collaborative filtering (BM3D)[12] is a state-of-the-art denoising method[13] with a substantial number of applications in the MRI domain[14-16]. However, these denoising methods' long calculation time and requirement for a carefully designed prior[17] have prevented their becoming established in clinical MRI settings.

Convolutional neural networks (CNNs)[18] use a series of convolution and non-linear activation operations, and thus show good flexibility and capability for learning a hierarchy of features, which makes them useful in image denoising. However, when a CNN goes to a certain depth, its gradient tends to vanish, causing a degradation in performance. To address this problem in deep CNNs, He et al. developed a residual learning (RL) approach that involves adding skip connections to a convolution block to connect low-level features directly to high-level representations[19]. RL allows for considerable increases in the depth and width of a network, and it has been shown to be efficient at image processing including classification[20], single image super-resolution[21], and denoising[22].

It is easier to predict the residual difference between the desired output and the input for a given stack of nonlinear layers in deep-learning models than to directly optimize the original mapping.[19] Many existing image denoising networks attempt to apply RL to approximate residual noise[22-28], based on the assumption that a noisy observation may be expressed as a combination of a clean image and noise. Zhang et al.[22] were the first to use RL with batch normalization (BN) in neural networks to generate denoising DnCNN, which are capable of blind Gaussian denoising of natural images. The application of DnCNNs in the MRI field has also achieved remarkable results, such as removing artificial Rician noise[23] and real noise in diffusion-weighted brain MR images[24]. In addition to DnCNN, various other RL-based denoising CNNs have been developed for MRI applications. Xie et al.[25] improved denoising performance in arterial spin labeling (ASL) MRI by preserving the spatial resolution of input images during model learning. Li et al[26] used Rician denoising to process synthetic and clinical brain MR images. RL combined with a fully convolutional network (FCN) by Ulas et al.[27], and an encoder-decoder structure by Tripathi et al.[28] were also used to denoise MR images.

Given the limited amount of available data, many MRI denoising networks have been trained using synthetic data with a given noise variance over an entire image. However, the noise distributions in actual MR images are often more complicated than those in these synthetic data-based approaches. For example, reconstruction methods such as sensitivity encoding (SENSE) and generalized autocalibrating partial parallel acquisition (GRAPPA) contribute to the non-stationarity of the noise variance. In this study, we explored deep learning-based methods for denoising 3D FSE MR images based on real noise distributions.

It is a common practice to increase the SNR (i.e., denoise) of MR images using time integration methods[29], such as by adjusting the number of excitations (NEX) or the number of signal averages (NSA). As SNR is calculated as the mean signal over the standard deviation of the noise[30], it is generally proportional to the square root of the NEX ($\sqrt{NEX}$). Although higher-quality images are obtained by increasing the NEX, a time penalty results. Thus, to restrict imaging time within reasonable ranges, typical NEX values need to be adjusted to obtain an acceptable tradeoff between time and image quality.[31]

In this study, we assumed that spatial-varying noise information can be extracted from 2-NEX acquisitions. This enabled us to use 2-NEX acquisitions to achieve denoising that generated a much higher SNR gain than that generated by the standard method, which uses the time integral of 2-NEX images. The scan time penalty was controlled by replacing a 1-NEX acquisition protocol with rapid low-SNR 2-NEX acquisitions with protocol optimization. We also developed a CNN model to make use of the inherent noise map embedded in these 2-NEX acquisitions, and thereby achieved significantly improved denoising performance compared to that without the CNN model. Overall, under the framework of RL designed for 2-NEX images, we demonstrated that our developed CNN model outperformed state-of-the-art denoising methods in the 3D FSE MRI of knee joints.

## 2. METHODS

### 2.1 Spatial-Varying Noise Distribution in 3D FSE MRI

It is commonly accepted that both the real and imaginary parts of the original signal from a single-coil MR acquisition are corrupted with uncorrelated zero-mean and equal-variance Gaussian noise in the frequency domain. After complex Fourier transformation, a linear and orthogonal transformation method, the Gaussian characteristics of noise in real and imaginary images, denoted by $N(0, \sigma_0{}^2)$, are preserved. Consequently, if the MR acquisition is repeated, a second complex-valued image is obtained; then, arithmetic can be performed on the two complex-valued images, as their signals can be assumed to be the same. Specifically, a complex-valued signal-strengthened map can be obtained by summing MR image data and a complex-valued noise map can be obtained by subtracting MR image data. The real and imaginary components of this signal-strengthened map and the noise map continue to exhibit a Gaussian distribution, denoted by $N(0, \sigma^2)$, where $\sigma = \sqrt{2}\sigma_0$.

The magnitude image of the signal-strengthened map therefore follows a Rician distribution and approximates a Gaussian distribution when the signal is sufficiently high.[32] In contrast, the magnitude image of the noise map follows a Rayleigh distribution, with the mean and variance given by Eq. (1), as follows.

$$\mu_R = \sigma\sqrt{\frac{\pi}{2}} \quad \text{and} \quad \sigma_R^2 = (2 - \frac{\pi}{2})\sigma^2 \tag{1}$$

Phased array coils with multiple coil elements, in which the complex Gaussian assumption of noise is valid in each coil in the frequency domain, are commonly used in MRI. If the $k$-space is fully sampled, the final composite magnitude image will follow a Rayleigh distribution in the background noise-only region, or a noncentral chi ($nc - \chi$) distribution, which take the signal intensities, but not the noise correlations, into consideration.[30]

However, in addition to the non-negligible correlations in phased array systems, the commonly employed subsampling acceleration technique in MRI also increases the complexity of noise distributions.[33] To illustrate this in 3D FSE MRI, we used an eight-channel knee coil with a SENSE acceleration factor of 2 to acquire 3D FSE images. Figure 1 shows two signal-strengthened magnitude images and the corresponding noise maps calculated using the aforementioned method. We generated the histograms of noise and fitted them with a Rayleigh distribution. If the noise follows an $nc - \chi$ distribution, its square will follow a noncentral chi-squared ($nc - \chi^2$) distribution.[34] It can be seen in Figure 1 that the noise showed a non-stationary pattern and did not exactly follow either a simple Rayleigh distribution or an $nc - \chi$ distribution. We also calculated the local variance of the noise maps in patches of $3 \times 3$ pixels (Figure 1), which revealed an obvious spatial variation in the noise variance. Thus, based on the assumption that noise is non-stationary in a given set of MR images, we considered that it would be more appropriate to use real MR images with a true noise distribution than simple synthetic noise to train a denoising network.

## 2.2 Network Architecture

A key aspect in the development of our proposed network was to use RL to learn the inherent real noise distribution in a dual-NEX acquisition, and thereby realize denoising of MR images that was superior to that obtainable with existing methods. The overall pipeline of our method, with some intermediate features of a typical slice, is illustrated in Figure 2.

The CNN consists of three modules: a feature extraction module, a bridge module, and an assembly module.

- Feature extraction module. The first six 128-kernel convolutional layers serve as the feature extraction module, which extracts certain low-level features (b) of the noise from the 2-NEX input (a). Our CNN is able to handle both dual-input (a) and single-input data (a').

- Bridge module. This module further optimizes the noise features using two parallel blocks: a transporting block and a residual block. The transporting block uses a 64-kernel convolutional layer with BN to maintain the flow of the original 2-NEX input information. The residual block provides a coarse evaluation of the denoised image by implementing the first stage of RL on the original 2-NEX inputs. An intermediate RL output (d) is then obtained by summing the first-stage residual difference (c) and a skip connection from the average of the 2-NEX input (a'). The residual filter feature maps (e) show features from the preliminary denoised intermediate RL output (d), while the transporting block feature maps (f) mostly inherit noise features from the feature extraction module.

- Assembly module. In this module, another convolutional layer is applied to consolidate the concatenated residual block feature maps (e) and transport the block feature maps (f). The remaining four convolution layers employ 64 kernels to generate high-level noise features. A sophisticated evaluation of the noise is then generated as the second residual difference (g). This residual difference and the previous coarsely denoised intermediate RL output (d) are summed to obtain the final denoised output (h).

## 2.3 Two-Step Residual Learning for 2-NEX Images

The noise map for 2-NEX images may be calculated as the difference between two individual NEX images, and thus we assumed that keeping each NEX image separated in the network would facilitate the determination of the inherent noise information. General RL models used for denoising requires the consistency of format between the input and the target output due to direct application of skip connections. Some variants of RL models, such as DL_ASL[25], does not have this limit after adding convolution to skip connections. These methods, however, are initially proposed for single input. In our network we designed RL to process two independent NEX images as input to use the inherent noise information in 2-NEX data for residual difference prediction and used their average for the skip connection.

We implemented RL in two stages in our network. The first stage of RL calculated a coarse residual difference and built its skip connection on the average of 2-NEX inputs, with its output denoted as the intermediate RL output. The second stage of RL generated a more sophisticated second-stage residual difference using features from both the 2-NEX input and the intermediate RL output. The final output was obtained from the second-stage residual difference with a skip connection to the intermediate RL output. This RL design ensured that the network made full use of both the strengthened signal and the inherent noise information from 2-NEX images.

## 2.4 Loss Function

The mean squared error, $l_2$, is probably the most widespread and convenient error measure used in image processing. However, it is widely accepted that $l_2$ does not correlate well with human perception of image quality.[35] The structural similarity index measure (SSIM)[36] is a commonly used reference-based index that evaluates images while accounting for the fact that the human visual system is sensitive to changes in local structure.[37] We thus combined $l_2$ loss and SSIM loss to generate an output with sharper edges and clearer details than either loss alone, as described in Eq. (2) below.

$$argmin_f||f(I_{2NEX}) - I_{8NEX})||_2^2 * (1 - SSIM(f(I_{2NEX}), I_{8NEX})) \qquad (2)$$

## 2.5 Evaluation Methods

The image denoising performance of our method was measured using the peak-signal-to-noise ratio (PSNR) and the SSIM value. The PSNR describes the denoising quality, and the SSIM is determined by modeling any image distortion as a combination of brightness, contrast, and structure correlation.[38]

Given a reference image $f$ and a test image $g$, with mean luminance $\mu_f$ and $\mu_g$, respectively, standard deviation $\sigma_f$ and $\sigma_g$, respectively, and covariance $\sigma_{fg}$ between f and g, the PSNR and SSIM between $f$ and $g$ are defined by Eqs (3) and (4), as follows:

$$PSNR(f,g) = 10log_{10}(\frac{255^2}{MSE(f,g)}) \qquad (3)$$

$$SSIM(f,g) = \frac{(2\mu_f\mu_g + C_1)(2\sigma_{fg} + C_2)}{(\mu_f^2 + \mu_g^2 + C_1)(\sigma_f^2 + \sigma_g^2 + C_2)}, \qquad (4)$$

where MSE is the mean square error between $f$ and $g$, and $C_1$ and $C_2$ are constants used to avoid division by zero. A higher PSNR indicates a higher image quality, and the closer that the SSIM values of two images are to 1, the more similar are the two structures. Matlab R2021a (Mathworks, Natick, MA, USA) was used for image analysis.

The quality of the patient datasets was independently reviewed for perceived SNR, overall image quality, and structure visibility by a radiologist with specialty fellowship training in musculoskeletal radiology. The key anatomical structures of the knee that were evaluated were the cartilage, anterior cruciate ligament, posterior cruciate ligament, medial collateral ligament, lateral collateral ligament, medial meniscus, lateral meniscus, extensor tendons, and bone.

2.6 Methods for Comparison and Training Settings

The performance of our method was evaluated and compared with the following state-of-the-art methods: BM3D[12], DnCNN[22], DL-ASL[25], and RicianNet[26]. For the 2-NEX input, the concatenation of two individual complex-valued images was denoted as dual-input, while the use of their average as the input was denoted as single-input. The real noise distribution was inherently present in the dual-input scenario but was absent from the single-input scenario. Because BM3D and DnCNN can only process single inputs, dual-input results were not obtained for these two methods. We also trained separated networks on three different imaging planes; i.e., the axial, coronal, and sagittal planes. We performed slice-by-slice denoising in 2D instead of direct 3D denoising for all denoising methods compared in this study, including our novel method.

Our CNN has 14 layers, and each convolution has filter size of $3 \times 3$, stride 1, and padding 1. In addition, it employs BN and the rectified linear unit (ReLU)[39] activation function. We used the Adam optimizer and the ReduceLROnPlateau monitor with an initial learning rate of 0.0001, which is decayed by a factor of 0.2 when the loss stops decreasing for 10 epochs. The entire 214 $\times$ 214 images were used as inputs, with a batch size of eight. All of the models were optimized using the loss function, as described in Eq. (2). As more kernels allow the CNN to capture more features[40,41], we doubled the number of convolution filters in each layer in the DnCNN (from 64 to 128) to enable a fair comparison of these methods.

2.7 Data Acquisition

The datasets were collected using a 3D proton density-weighted FSE/TSE VISTA™ pulse sequence on a Philips Achieva TX 3.0T MRI instrument (Philips Healthcare, Best, Netherlands) with an eight-channel receiver knee coil (Invivo, Gainesville, FL, USA). All MRI examinations were conducted under the approval of the Institutional Review Board. The MRI parameters were as follows: a repetition time/echo time of 900/33.6 ms; 150 slices with an isotropic resolution of $0.8 \times 0.8 \times 0.8$ mm; an echo train length of 42; and a SENSE acceleration factor of 2. The imaging acquisition time per NEX was approximately 2.9 min. The 2-NEX acquisition provided the input of the network, and we collected high-SNR images (which served as the target images) from the 8-NEX acquisition. Datasets including 8-NEX 3D FSE MRI data were collected from 67 healthy volunteers. 50 of these datasets were used for training and 17 were used for testing. An additional 40 3D FSE MRI datasets with 2-NEX were collected from 40 patients (categorized into the four Kellegren and Lawrence (KL) grades for the classification of osteoarthritis, with KL4 being the most severe) exhibiting various stages of osteoarthritis, and these datasets were used for testing.

## 3. Results

3.1 Network Performance with Single-Input and Dual-Input Data

Table 1 shows the average PSNR and SSIM results for the sagittal plane testing of healthy volunteer datasets. We denoted 2NEX-avg as the standard results after averaging two complex-valued images of 2-NEX acquisitions. The four deep-learning models were superior to the traditional denoising method, BM3D, and our novel model using dual-input images achieved the best performance.

Although the single-input and dual-input training data were from the same batch, with equal acquisition times, the denoising performance with the dual-input data was always superior to the single-input data. This was because our RL method for 2-NEX images was designed to take advantage of the noise information from 2-NEX images, which enabled our model to process dual-input data with better denoising ability than single-input data.

Figure 3 shows the typical denoising results from the test datasets and the corresponding 8-NEX ground truth. Use of the dual-input data resulted greater similarity to the 8-NEX ground truth than use of the single-input data. Of all the methods tested, our model with dual-input data suppressed noise most uniformly, even in regions with high noise levels. Facilitated by the separation and concatenation operations of the parallel transporting and residual blocks, the RL design for 2-NEX images ensured that the final output used both original and residual features.

3.2 Network Performance for Denoising in 3D Datasets

Table 2 shows the mean PSNR and SSIM denoising results from healthy volunteer testing datasets in the axial, coronal, and sagittal planes. Figure 4 shows the result of denoising the data of one healthy test subject in these three planes. The noise level and the distribution of 2NEX-avg images differed in each plane, even for the same knee, which is consistent with the spatially variant characteristics of noise. As we trained our model using MR images containing real noise, our method achieved denoising performance in all three planes that was superior to that achieved by standard averaging methods. Moreover, our denoising method clearly denoised MR images of critical structures, including bone, cartilage, menisci, and the anterior cruciate ligament.

The patient cohort group comprised 40 patients with varying severities of osteoarthritis of the knee. The implementation of our denoising approach afforded MR images of all anatomical structures that had superior perceived SNRs and improved overall image quality relative to standard averaging method. A typical result for a patient test dataset is shown in Figure 5. This subject had KL4 osteoarthritis, and the characteristic radiological findings seen in osteoarthritis, such as cartilage thinning, osteophytosis, bone marrow cysts, bone marrow edema, joint effusions, and para-labral cysts, were all clearly identifiable following noise suppression using our method.

The performance of our method using data from both healthy volunteer and osteoarthritis patient datasets implies that our CNN is able to simultaneously denoise and retain anatomical structures. The beneficial reduction in overall patient scan time in three planes combined with improved image quality highlights its potential clinical utility.

### 3.3 Ablation Study

The feature extraction module and the assembly module constitute the nonremovable skeleton in our CNN. Thus, ablation studies were used to investigate the design of the bridge module. The modified network structure and the corresponding denoising results on the same test datasets are shown in Figure 6. Model-Tra represents our CNN with only a transporting block in the bridge module, and Model-Res corresponds to the CNN with only a residual block. Table 3 shows the mean PSNR and SSIM values of the three models for the healthy volunteer test datasets. Our model showed increased denoising performance compared to both the Model-Tra and Model-Res models. In particular, the transporting block preserved the overall noise information from the original 2-NEX input, and the residual block extracted more subtle noise information from the already coarsely denoised intermediate residual output. The parallel block structure thus provided sufficient information for our novel CNN, contributing to its improved performance in both PSNR and SSIM metrics.

## 4. DISCUSSION

3D FSE shows clinical potential for MRI-based assessments of articular cartilage, menisci, ligaments, tendons, and nerves.[42,43] To further utilize 3D FSE high-through-plane resolution to visualize complex anatomical structures, we developed a method for 3D FSE knee joint MR image denoising.

Prior knowledge of a noise distribution facilitates denoising; however, real noise distributions of typical 3D MR images often have spatial variations and are difficult to characterize. Li et al.[44] obtained a 3D FSE MRI noise map using additional noise-only acquisition with radiofrequency (RF) excitations switched off and by following a routine reconstruction pipeline. Instead, we used 2-NEX acquisition to provide a network with inherent noise information. Our CNN was specifically designed for dual-input data and it contains feature extraction, bridge, and assembly modules to extract, integrate, and transfer the features of the 2-NEX input. Assisted by the L2 norm and SSIM loss, the two-step RL parallel transporting and residual block design of our CNN ensured that it comprehensively learned real noise information from 2-NEX training data. The experimental results suggest that our method was superior to state-of-the-art denoising networks. Specifically, the overall improvement in the quality of 3D FSE MR images of osteoarthritis patients' knees processed by our method compared with the quality of images processed with standard averaging methods demonstrates the clinical potential of our approach.

FSE acquisition plays a central role in clinical MRI and is used for various anatomical structures. In this study, we focused on denoising 3D FSE MR images of knee joints. This method may be extended to 3D FSE imaging of other anatomical structures and imaging applications based on 3D FSE acquisitions. Moreover, quantitative analysis based on 3D FSE may also benefit from image denoising. Magnetization-prepared 3D FSE has been used for fast 3D quantitative parametric MRI[45,46] due to its high SNR efficiency and insensitivity to T1 relaxation effects during FSE readout under the Carr–Purcell–Meiboom–Gill condition. As accurate MRI quantification depends on the noise characteristics in quantification models, our method is potentially beneficial to quantitative MRI based on 3D FSE acquisitions. In addition, our deep-learning denoising CNN is likely to be generalizable to other MRI pulse sequences using the multi-NEX acquisition method, such as 2D FSE.

For protocols that use multi-NEX during FSE acquisition, there is no scan time penalty when applying our method for denoising. For protocols with 1-NEX acquisition, a 2-NEX scan of simply two repeated 1-NEX scans doubles the scan time. For 3D FSE, there is the flexibility of adjusting the pulse sequence parameters to achieve a tradeoff between scan time and SNR. The scan time can be reduced by increasing the echo train length (i.e., the number of refocusing RF pulses in one train). However, increasing the echo train may also increase image blurring. The flip angle train can be designed (i.e., by reducing the minimum refocusing flip angle) to reduce blurring. The train duration can also be reduced by increasing the readout bandwidth, thus reducing image blurring resulting from the increased echo train length. In addition, the scan time can be reduced by performing greater undersampling of the $k$-space. However, the scan time reduction for such a protocol adjustment comes at the cost of the SNR. Thus, the scan time penalty of a 2-NEX scan involves a tradeoff with an SNR penalty. A final denoised image obtained by our method therefore has a much higher SNR than the sum of the 2-NEX acquisitions which is expected to have SNR similar to 1-NEX acquisition with comparable scan time.

There are several limitations to this study. First, our method was a supervised approach. Thus, if the imaging protocol is changed, resulting in a substantial change in the real noise distribution, the network may need to be retrained with target images acquired using the corresponding protocol. It is thus important to investigate approaches that require a few target images to train the network, under the same principle proposed in this work; namely, using the inherent real noise in a dual-input complex-valued image. Second, our training data were all from healthy subjects, and thus our method may be relatively weak at processing real patient data. That is, except for the reconstruction methods, dielectric and inductive losses in a sample may also influence the noise distribution[47]. Tissues with different water content may contribute differently to noise.[48] This emphasizes the need to use real MR images of different anatomical structures and inhomogeneous dielectric properties for network training. Third, we used 8-NEX 3D FSE images as the ground truth for training. The acquisition of these images requires a long scan time, and thus it should be investigated whether images obtainable via other, shorter acquisitions can be used as the ground truth. Fourth, we performed denoising slice-by-slice using a 2D network. Given the redundancy of 3D data, it should be explored whether our CNN can be extend to 3D complex-valued CNN[49], as this would enable exploration of the 3D noise features of complex-valued MRI data and the development of more convenient applications.

## 5. CONCLUSION

We developed an RL-based CNN that uses the spatial-variant noise information from dual-NEX acquisition to denoise MR images. This approach achieved significantly improved SNR gain compared to standard methods of averaging multi-NEX acquisition. Our method also outperformed existing methods in the denoising of 3D FSE MR images of knee joints. The results from patients exhibiting various stages of osteoarthritis show that our method has potential clinical utility for knee joint MR imaging and provides a new perspective for MRI denoising tasks.

## ACKNOWLEDGEMENT

This study was supported by a grant from the Innovation and Technology Commission of the Hong Kong SAR (Project MRP/001/18X), a grant from the Faculty Innovation Aware, the Chinese

University of Hong Kong, and a grant from the Research Grants Council of the Hong Kong SAR (Project SEG CUHK02).

## Reference

1.  Scuderi GR, Tria AJ. The knee: a comprehensive review. 2010.
2.  Leung DG, Carrino JA, Wagner KR, Jacobs MA. Whole-body magnetic resonance imaging evaluation of facioscapulohumeral muscular dystrophy. *Muscle & nerve*. 2015,52(4):512-520. http://dx.doi.org/10.1002/mus.24569
3.  RT YH, Sasaki M, RT KE, RT HG. Hitachi's Prime Fast Spin Echo Technology: Efficacies in Improving Image Quality and Usability. 2008.
4.  Hennig J, Nauerth A, Friedburg H. RARE imaging: a fast imaging method for clinical MR. *Magnetic resonance in medicine.* 1986,3(6):823-833. http://dx.doi.org/10.1002/mrm.1910030602
5.  Kijowski R, Davis KW, Woods MA, et al. Knee joint: comprehensive assessment with 3D isotropic resolution fast spin-echo MR imaging—diagnostic performance compared with that of conventional MR imaging at 3.0 T. *Radiology*. 2009,252(2): 486-495. http://dx.doi.org/10.1148/radiol.2523090028
6.  Li CQ, Chen W, Rosenberg JK, et al. Optimizing isotropic three-dimensional fast spin-echo methods for imaging the knee. *Journal of Magnetic Resonance Imaging*. 2014,39(6):1417-1425. http://dx.doi.org/10.1002/jmri.24315
7.  Alsop DC. The sensitivity of low flip angle RARE imaging. *Magnetic resonance in medicine*. 1997,37(2):176-184. http://dx.doi.org/10.1002/mrm.1910370206
8.  Tomasi C, Manduchi R. Bilateral filtering for gray and color images. *Sixth international conference on computer vision (IEEE Cat. No. 98CH36271)*. *IEEE.* 1998:839-846. http://dx.doi.org/10.1109/ICCV.1998.710815

9.  Rudin LI, Osher S, Fatemi E. Nonlinear total variation based noise removal algorithms. *Physica D: nonlinear phenomena*. 1992,60(1-4):259-268. http://dx.doi.org/10.1016/0167-2789(92)90242-F

10. Buades A, Coll B, Morel JM. A non-local algorithm for image denoising. *2005 IEEE Computer Society Conference on Computer Vision and Pattern Recognition (CVPR'05). IEEE.* 2005,2:60-65. http://dx.doi.org/10.1109/CVPR.2005.38

11. Aharon M, Elad M, Bruckstein A. K-SVD: An algorithm for designing overcomplete dictionaries for sparse representation. *IEEE Transactions on signal processing*. 2006,54(11):4311-4322. http://dx.doi.org/10.1109/TSP.2006.881199

12. Dabov K, Foi A, Katkovnik V, Egiazarian K. Image denoising by sparse 3-D transform-domain collaborative filtering. IEEE Transactions on image processing. 2007,16(8):2080-2095. http://dx.doi.org/10.1109/TIP.2007.901238

13. Burger HC, Schuler CJ, Harmeling S. Image denoising: can plain neural networks compete with BM3D? *2012 IEEE conference on computer vision and pattern recognition*, Providence, RI: IEEE; 2012: 2392- 2399. http://dx.doi.org/10.1109/CVPR.2012.6247952

14. Lin XB, Qiu TS. Denoise MRI images using sparse 3D transformation domain collaborative filtering. *2011 4th International Conference on Biomedical Engineering and Informatics (BMEI). IEEE.* 2011,1:233-236. http://dx.doi.org/10.1109/BMEI.2011.6098330

15. Maggioni M, Katkovnik V, Egiazarian K, Foi A. Nonlocal transform-domain filter for volumetric data denoising and reconstruction. *IEEE transactions on image processing*. 2012,22(1):119-133. http://dx.doi.org/10.1109/TIP.2012.2210725

16. Foi A. Noise estimation and removal in MR imaging: The variance-stabilization approach. *2011 IEEE International symposium on biomedical imaging: from nano to macro. IEEE.* 2011:1809-1814. http://dx.doi.org/10.1109/ISBI.2011.5872758

17. Chang Y, Yan L, Chen M, Fang H, Zhong S. Two-stage convolutional neural network for medical noise removal via image decomposition. *IEEE Transactions on Instrumentation and Measurement*. 2019,69(6):2707-2721. http://dx.doi.org/10.1109/TIM.2019.2925881

18. Krizhevsky A, Sutskever I, Hinton GE. Imagenet classification with deep convolutional neural networks. *Advances in neural information processing systems*. 2012,25:1097-1105. http://dx.doi.org/10.1145/3065386

19. He K, Zhang X, Ren S, Sun J. Deep residual learning for image recognition. *Proceedings of the IEEE conference on computer vision and pattern recognition*. 2016:770-778. http://dx.doi.org/10.1109/CVPR.2016.90

20. Lee H, Kwon H. Going deeper with contextual CNN for hyperspectral image classification. *IEEE Transactions on Image Processing*. 2017,26(10):4843-4855. http://dx.doi.org/10.1109/TIP.2017.2725580

21. Kim J, Lee JK, Lee KM. Accurate image super-resolution using very deep convolutional networks. *Proceedings of the IEEE conference on computer vision and pattern recognition*. 2016:1646-1654. http://dx.doi.org/10.1109/CVPR.2016.182

22. Zhang K, Zuo W, Chen Y, Meng D, Zhang L. Beyond a gaussian denoiser: Residual learning of deep cnn for image denoising. *IEEE transactions on image processing.* 2017,26(7):3142-3155. http://dx.doi.org/10.1109/TIP.2017.2662206

23. Jiang D, Dou W, Vosters L, Xu X, Sun Y, Tan T. Denoising of 3D magnetic resonance images with multi-channel residual learning of convolutional neural network. *Japanese Journal of Radiology*. 2017:566-574. http://dx.doi.org/10.1007/s11604-018-0758-8

24. Kawamura M, Tamada D, Funayama S, et al. Accelerated acquisition of high-resolution diffusion-weighted imaging of the brain with a multi-shot echo-planar sequence: deep-learning-based denoising. *Magnetic Resonance in Medical Sciences*. 2021,20(1):99. http://dx.doi.org/10.2463/mrms.tn.2019-0081

25. Xie D, Li Y, Yang H, et al. Denoising arterial spin labeling perfusion MRI with deep machine learning. *Magnetic resonance imaging.* 2020,68:95-105. http://dx.doi.org/10.1016/j.mri.2020.01.005

26. Li S, Zhou J, Liang D, Liu Q. MRI denoising using progressively distribution-based neural network. *Magnetic resonance imaging*. 2020,71:55-68. http://dx.doi.org/10.1016/j.mri.2020.04.006

27. Ulas C, Tetteh G, Kaczmarz S, Preibisch C, Menze BH. DeepASL: Kinetic model incorporated loss for denoising arterial spin labeled MRI via deep residual learning. *International conference on medical image computing and computer-assisted intervention*. 2018:30-38. http://dx.doi.org/10.1007/978-3-030-00928-1_4

28. Tripathi PC, Bag S. CNN-DMRI: a convolutional neural network for denoising of magnetic resonance images. *Pattern Recognition Letters*. 2020,135:57-63. http://dx.doi.org/10.1016/j.patrec.2020.03.036

29. Sijbers J, Scheunders P, Bonnet N, Van Dyck D, Raman E. Quantification and improvement of the signal-to-noise ratio in a magnetic resonance image acquisition procedure. *Magnetic resonance imaging*. 1996,14(10):1157-1163. http://dx.doi.org/10.1016/S0730-725X(96)00219-6

30. Constantinides CD, Ergin A, Elliot RM. Signal-to-noise measurements in magnitude images from NMR phased arrays. *Magnetic Resonance in Medicine*. 1997,38(5):852-857. http://dx.doi.org/10.1109/IEMBS.1997.754578

31. Ting YN, Yuan HS, Chang HH, Chu WC. Automatic noise removal and effect of nEX setting on magnetic resonance images. *2011 4th International Conference on Biomedical Engineering and Informatics (BMEI). IEEE.* 2011,1:160-164. http://dx.doi.org/10.1109/BMEI.2011.6098280

32. Gudbjartsson H, Samuel P. The Rician distribution of noisy MRI data. *Magnetic resonance in medicine.* 1995,34(6):910-914. http://dx.doi.org/10.1002/mrm.1910340618

33. Aja-Fernández S, Vegas-Sánchez-Ferrero G, Tristán-Vega A. Noise estimation in parallel MRI: GRAPPA and SENSE. *Magnetic resonance imaging*. 2014,32(3):281-290. http://dx.doi.org/10.1016/j.mri.2013.12.001

34. Aja-Fernández S, Tristán-Vega A, Hoge WS. Statistical noise analysis in GRAPPA using a parametrized noncentral Chi approximation model. *Magnetic resonance in medicine*. 2011,65(4):1195-1206. http://dx.doi.org/10.1002/mrm.22701

35. Zhang L, Zhang L, Mou X, Zhang D. A comprehensive evaluation of full reference image quality assessment algorithms. *2012 19th IEEE International Conference on Image Processing. IEEE.* 2012:1477-1480. http://dx.doi.org/10.1109/ICIP.2012.6467150

36. Wang Z, Bovik AC, Sheikh HR, Simoncelli EP. Image quality assessment: from error visibility to structural similarity. *IEEE transactions on image processing.* 2004,13(4):600-612. http://dx.doi.org/10.1109/TIP.2003.819861

37. Zhao H, Gallo O, Frosio I, Kautz J. Loss functions for image restoration with neural networks. *IEEE Transactions on computational imaging*, 2016,*3*(1):47-57. http://dx.doi.org/10.1109/TCI.2016.2644865

38. Horé A, Ziou D. Image Quality Metrics: PSNR vs. SSIM. *2010 20th International Conference on Pattern Recognition.* 2010:2366-2369. http://dx.doi.org/10.1109/ICPR.2010.579

39. Krizhevsky A, Sutskever I, Hinton GE. Imagenet classification with deep convolutional neural networks. *Advances in neural information processing systems* 2012,25:1097-1105. http://dx.doi.org/10.1145/3065386

40. Chung HY, Chung YL, Tsai WF. An Efficient Hand Gesture Recognition System Based on Deep CNN. *2019 IEEE International Conference on Industrial Technology (ICIT)*. 2019:853-858. http://dx.doi.org/10.1109/ICIT.2019.8755038

41. Zhang Y, Lu W, Ou W, et al. Chinese medical question answer selection via hybrid models based on CNN and GRU. *Multimedia Tools and Applications.* 2020,79(21):14751–14776. http://dx.doi.org/10.1007/s11042-019-7240-1

42. Altahawi F, Pierce J, Aslan M, Li X, Winalski CS, Subhas N. 3D MRI of the Knee. *Seminars in Musculoskeletal Radiology. Thieme Medical Publishers, Inc.* 2021, 25(03):455-467. http://dx.doi.org/10.1055/s-0041-1730400

43. Notohamiprodjo M, Horng A, Kuschel B, et al. 3D-imaging of the knee with an optimized 3D-FSE-sequence and a 15-channel knee-coil. *European journal of radiology.* 2012,81(11):3441-3449. http://dx.doi.org/10.1016/j.ejrad.2012.04.020

44. Li CQ, Chen W, Rosenberg JK, et al. Optimizing isotropic three-dimensional fast spin-echo methods for imaging the knee. *Journal of Magnetic Resonance Imaging*. 2014,39(6):1417-1425. http://dx.doi.org/10.1002/jmri.24315

45. Chen W, Takahashi A, Han E. 3D quantitative imaging of T1rho and T2. *Proceedings of the 19th Annual Meeting of ISMRM.* 2011,231.

46. Jordan CD, McWalter EJ, Monu UD, et al. Variability of CubeQuant T1ρ, quantitative DESS T2, and cones sodium MRI in knee cartilage. *Osteoarthritis and cartilage.* 2014;22(10):1559-1567. http://dx.doi.org/10.1016/j.joca.2014.06.001

47. McVeigh ER, Henkelman RM, Bronskill MJ. Noise and filtration in magnetic resonance imaging. *Medical physics.* 1985,12(5):586-591. http://dx.doi.org/10.1118/1.595679

48. Hoult DI, Lauterbur PC. The sensitivity of the zeugmatographic experiment involving human samples. *Journal of Magnetic Resonance*. (1969),1979,34(2):425-433. http://dx.doi.org/10.1016/0022-2364(79)90019-2

49. Zhang Z, Wang H, Xu F, Jin YQ. Complex-valued convolutional neural network and its application in polarimetric SAR image classification. *IEEE Transactions on Geoscience and Remote Sensing.* 2017,55(12):7177-7188. http://dx.doi.org/10.1109/TGRS.2017.2743222

**Figures**

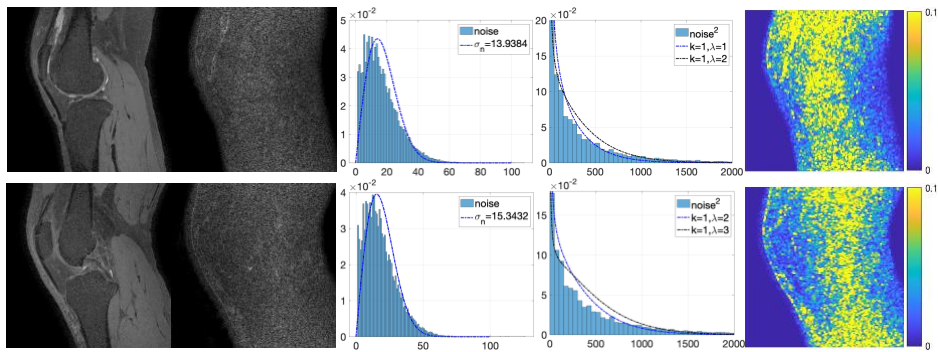

Figure 1: Two typical slices of 2-NEX images. From left to right: signal-strengthened images, noise maps, histograms of the noise with corresponding fitted Rayleigh probability density functions, histograms of the noise squared with noncentral chi-squared distribution probability density functions, and local variance maps of noise.

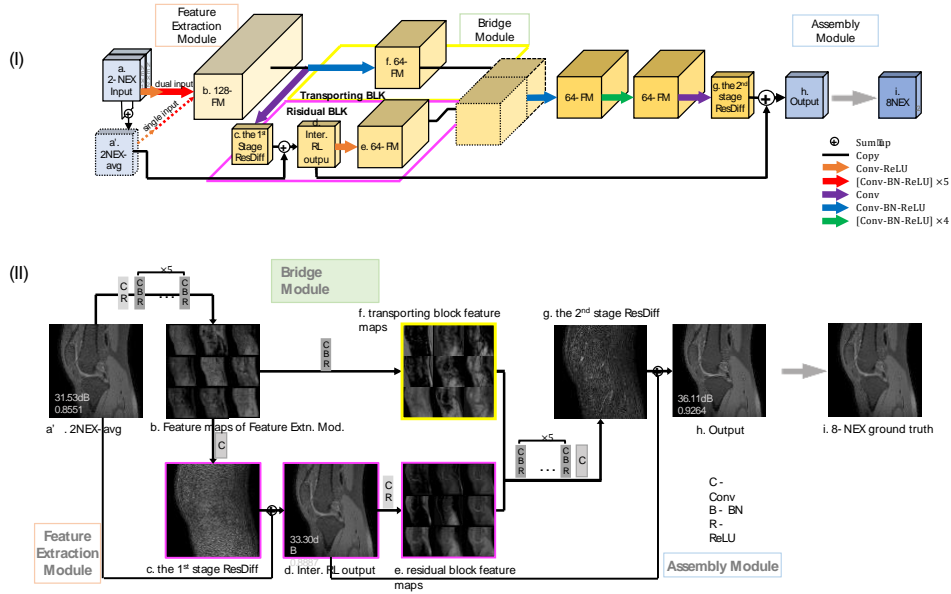

Figure 2: Our novel convolutional neural network (CNN) architecture (I) and an illustration of some of the channel features at some layers of a typical slice (II). Our CNN can manage both dual-input (solid line (a)) and single-input (dashed line (a')) data.

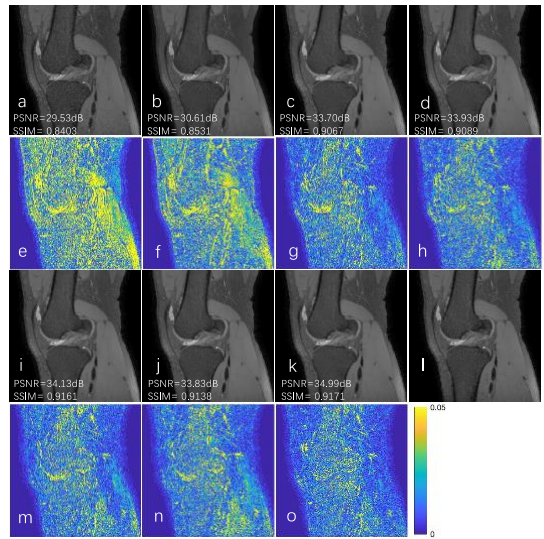

Figure 3: Denoising results of a typical slice of the volunteer test dataset. The first row shows (a) the single-input 2NEX-avg and its corresponding denoised output using (b) BM3D, (c) DnCNN, and (d) our model. The third row shows the denoised outputs from a dual-input 2-NEX using (i) DL_ASL, (j) RicianNet, (k) our model, and (l) the ground truth. The second and fourth rows are residual differences between corresponding images and the 8-NEX ground truth. 2NEX-avg = the standard results after averaging two complex-valued images of 2-NEX acquisitions.

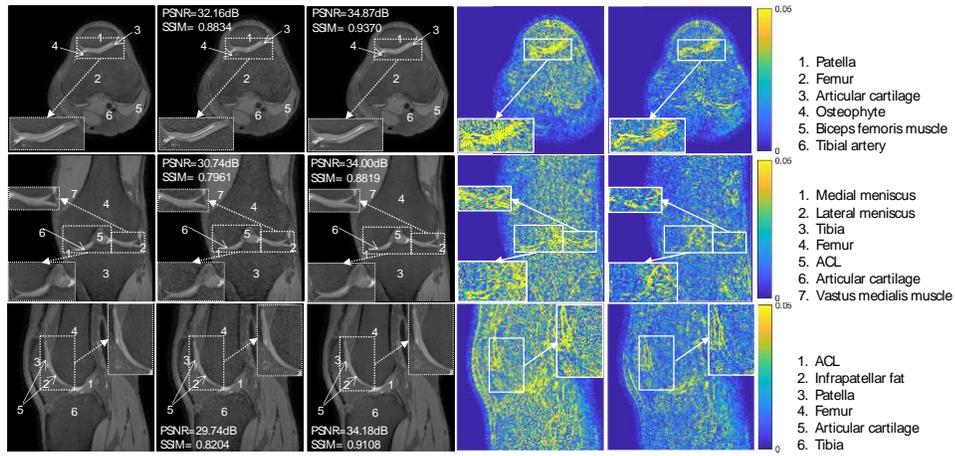

Figure 4: Results of denoising typical slices from different planes of the volunteer test dataset. From top to bottom row: the axial, coronal, and sagittal planes. From left to right: a ground truth 8-NEX image, an input 2NEX-avg image, an image denoised using our model, the residual difference between the 2NEX-avg input and the ground truth, and the residual difference between the denoised image and the ground truth. 2NEX-avg = the standard results after averaging two complex-valued images of 2-NEX acquisitions.

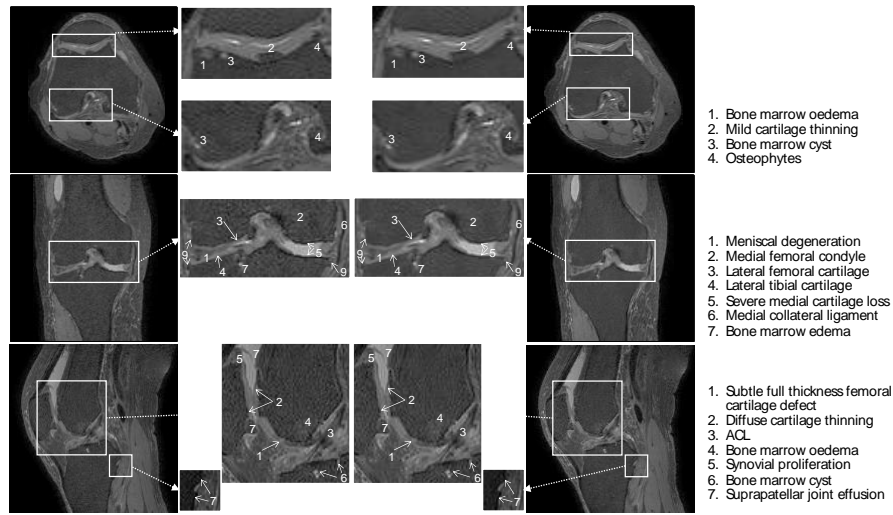

Figure 5: Results of denoising typical slices from different planes of an osteoarthritis patient test dataset. From top to bottom row: the axial, coronal, and sagittal planes. From left to right: an input 2NEX-avg image and the image denoised using our model. The image in the white box is enlarged to highlight the fine structural information. 2NEX-avg = the standard results after averaging two complex-valued images of 2-NEX acquisitions.

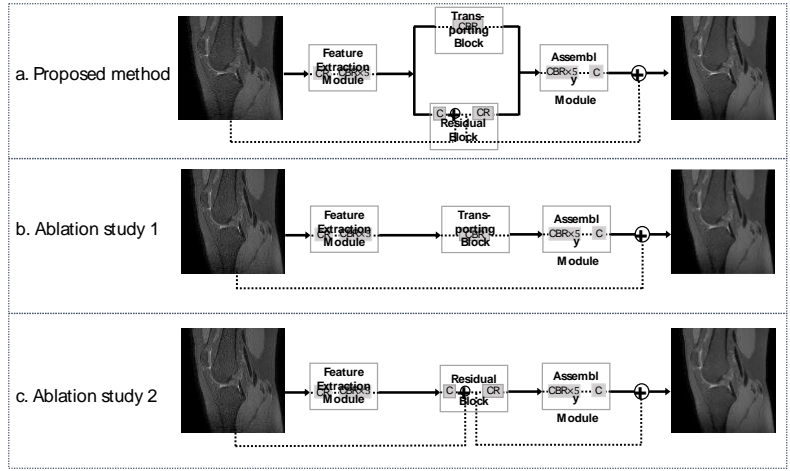

Figure 6: Our convolutional neural network (a) and its variants, Model-Tra (b) and Model-Res (c).

Tables

| | PSNR | | SSIM | |
|---|---|---|---|---|
| | Single-input | Dual-input | Single-input | Dual-input |
| 2NEX-avg | 31.4114±2.2761 | | 0.86612±0.03834 | |
| BM3D* | 32.2920±2.3124 | – | 0.88810±0.03103 | – |
| DnCNN* | 34.0181±2.2646 | – | 0.91976±0.02026 | – |
| DL_ASL | 34.0692±2.2068 | 34.4133±2.1456 | 0.92052±0.02125 | 0.92367±0.02047 |
| RicianNet | 34.0455±2.2154 | 34.5632±2.2042 | **0.92055±0.02066** | 0.92388±0.02031 |
| Our Model | **34.1432±2.2244** | <u>**34.7233±2.2108**</u> | 0.92054±0.02123 | <u>**0.92459±0.02047**</u> |

* BM3D and DnCNN only process single input.

Table 1. Mean PSNR and SSIM values indicating the performance of denoising methods in the sagittal plane. The BM3D and DnCNN methods only process single inputs.

| | PSNR | | SSIM | |
| | 2NEX-avg | Our Model | 2NEX-avg | Our Model |
|---|---|---|---|---|
| axial | 31.1043±2.6417 | 33.4544±2.4321 | 0.86473±0.05759 | 0.91581±0.03581 |
| coronal | 33.2752±2.2388 | 35.8279±2.3827 | 0.91528±0.02548 | 0.94977±0.01524 |
| sagittal | 31.4160±2.2530 | 34.7236±2.1836 | 0.86614±0.03823 | 0.92463±0.02046 |

Table 2. Mean PSNR and SSIM values of the test datasets denoised in the axial, coronal, and sagittal planes.

| | PSNR | SSIM |
|---|---|---|
| Model | 34.7233±2.2108 | 0.92459±0.02047 |
| Model-Tra | 34.5816±2.2452 | 0.92308±0.02157 |
| Model-Res | 34.6493±2.1935 | 0.92320±0.02115 |

Table 3. Mean PSNR and SSIM values of the ablation study models used to test the performance of the models in the sagittal plane.